\documentclass[a4paper]{article}

\usepackage{INTERSPEECH2021}
\usepackage{amsmath,graphicx,booktabs,amssymb}
\usepackage{makecell,multirow,bm}
\newcommand{\tabincell}[2]{\begin{tabular}{@{}#1@{}}#2\end{tabular}}

\usepackage{algorithm}
\usepackage{algorithmic} 
\title{Improved Meta-learning Training for Speaker Verification}
\name{Yafeng Chen, Wu Guo, Bin Gu}
\address{
  National Engineering Laboratory for Speech and Language Information Processing,
University of Science and Technology of China, Hefei, China}
\email{\{yfchen97, bin2801\}@mail.ustc.edu.cn, guowu@ustc.edu.cn}

\begin{document}

\maketitle
\begin{abstract}
Meta-learning (ML) has  recently become a research hotspot in speaker verification (SV). We introduce two methods to improve the meta-learning training for SV in this paper. For the first method, a backbone embedding network is first jointly trained with the conventional cross entropy loss and prototypical networks (PN) loss. Then, inspired by speaker adaptive training in speech recognition, additional transformation coefficients are trained with only the PN loss. The transformation coefficients are used to modify the original backbone embedding network in the x-vector extraction process. Furthermore, the random erasing (RE) data augmentation technique is applied to all support samples in each episode to construct positive pairs, and a contrastive loss between the augmented and the original support samples is added to the objective in model training. Experiments are carried out on the Speaker in the Wild (SITW) and VOiCES databases. Both of the methods can obtain consistent improvements over existing meta-learning training frameworks. By combining these two methods, we can observe further improvements on these two databases.
\end{abstract}
\noindent\textbf{Index Terms}: speaker verification, meta-learning, prototypical networks, random erasing, contrastive learning

\section{Introduction}

Over the years, the combination of i-vectors \cite{dehak2010front,rohdin2018end} and probabilistic linear discriminant analysis (PLDA) \cite{garcia2011analysis} has been the dominant approach for speaker verification (SV) tasks. With the great success of deep neural networks (DNNs) in machine learning, more efforts have been focused on how to learn DNN-based speaker representations, known as speaker embeddings. Many novel DNN embedding-based systems have been proposed, and they have achieved comparable or even superior performance to the traditional i-vector paradigm. The most representative system is the x-vector framework \cite{snyder2018x}.

For most of these DNN embedding systems, the networks are trained to classify speakers with the cross entropy (CE) loss, which aims to maximize the between-class variance in the training set. However, SV is to verify the claimed identity of a person for a given speech, which aims to maximize the between-class variance while minimizing the within-class variance of the trials. If DNN systems take this into consideration in model training, the speaker embeddings can be more discriminative. To address this problem, there are several new approaches \cite{ko2020prototypical, kye2020meta, wang2019centroid} in meta-learning that try to learn a shared metric space between the embeddings of the test examples and the known classes. In \cite{ko2020prototypical}, Tom Ko et al. adopted prototypical networks (PN), a typical meta-learning architecture, to boost the discriminative power for speaker embedding extraction. However, optimizing only for the classes within each episode may be insufficient for learning discriminative embeddings for unseen classes in PN training, so \cite{kye2020meta} further classifies every sample in each episode against the whole training class (referred to as global classification). Impressive improvements can be observed in short duration speaker recognition.

As mentioned in \cite{kye2020meta}, the global classification loss, such as the cross entropy (CE), plays an important role in the SV task. The most important question is to balance the conventional PN loss and the global classification loss for the SV task. In this work, we present a novel Meta-Learning method with additional Transformation Coefficients (MLTC), which is similar to the adaptive training method in speech recognition. For the proposed method, a backbone embedding network is first jointly trained with the PN loss and CE loss. This step is the same as the training process in \cite{kye2020meta}. The parameters of the trained model are fixed, and then additional transformation coefficients are trained with only the PN loss. In the x-vector extraction process, the transformation coefficients modify the original backbone embedding network to obtain embeddings.

Prototypical networks construct positive- (from the same speaker) and negative- (from different speakers) pairs from the support and query sets for model training in each episode. These two sets are randomly divided among all samples in each minibatch, which may induce extra noise in model training. To solve this problem, we apply the random erasing (RE) data augmentation technique \cite{zhong2020random} to all support samples and construct positive pairs between the original and augmented samples. Then, an addition contrastive loss (ACL), which is calculated between the augmented and the original support set, is added to the training objective. The above method is similar to contrastive learning \cite{he2020momentum, chen2020improved, chen2020simple},which is widely used in unsupervised learning.

We validate our proposed methods on the Speaker in the Wild (SITW) \cite{mclaren2016speakers} and VOiCES \cite{nandwana2019voices} datasets. The experimental results show that MLTC can achieve superior performance to existing meta-learning training frameworks, and the combination with ACL can obtain further improvements. 

The remainder of this paper is organized as follows. In Section 2, we briefly introduce the related works about the meta-learning method in SV tasks. In Section 3, we describe the improved meta-learning training methods in detail. The experimental setup, the results and analysis are presented in Section 4. Finally, conclusions are given in Section 5.

\section{Related works}

\subsection{Prototypical networks}

Prototypical networks train a model episodically; each episode is composed of one minibatch containing $N$ speakers. In each episode, $S=\{S_1, ..., S_N\}$ is the support set and $Q=\{Q_1, ..., Q_N\}$ is the query set. Let $S_n=\{(\textbf{x}_{n,1}^s, y_n^s), ..., (\textbf{x}_{n,|S_n|}^s, y_n^s)\}$ and $Q_n=\{(\textbf{x}_{n,1}^q, y_n^q), ..., (\textbf{x}_{n,|Q_n|}^q, y_n^q)\}$ denote the sets of samples labeled with speaker $n$, where $|S_n|$ and $|Q_n|$ stand for the number of samples in $S_n$ and $Q_n$, respectively. $\textbf{x}_{n,i}^s$ and $\textbf{x}_{n,i}^q$ are $d$-dimensional acoustic feature vectors of the samples, and $y_n^s \in \{1, ..., N \}$ and $y_n^q \in \{1, ..., N \}$ are the corresponding labels. The centroid of each speaker $n$ is calculated as the mean of the embeddings in the support set.
\begin{equation}
  \textbf{c}_n = \frac{1}{|S_n|}\sum\limits_{(\textbf{x}_{n,i}^s, y_n^s)\in S_n}f_\theta(\textbf{x}_{n,i}^s)
\end{equation}
where $f_\theta(\cdot)$ is an embedding extractor with parameters $\theta$.
Then, the probability of each query sample $(\textbf{x}_{n,i}^q, y_n^q) \in Q_n$ belonging to speaker $n$ is computed as follows. 
\begin{equation}
  p(y_n^q|\textbf{x}_{n,i}^q,S;\theta) = \frac{exp(-d(f_\theta(\textbf{x}_{n,i}^q), \textbf{c}_n))}{\sum\limits_{n^{\prime}=1}^N exp(-d(f_\theta(\textbf{x}_{n,i}^q),\textbf{c}_{n^{\prime}}))}
\end{equation}
The cosine and squared Euclidean distances are commonly adopted as metric functions $d(\cdot)$ in Eq. 2. Furthermore, the PN loss for each episode is calculated.
\begin{equation}
\begin{split}
  L_{PN} = -\frac{1}{N}\sum\limits_{n=1}^N\frac{1}{Q_n}\sum\limits_{(\textbf{x}_{n,i}^q, y_n^q) \in Q_n}log  p(y_n^q|\textbf{x}_{n,i}^q,S;\theta)
\end{split}
\end{equation}
The PN loss is always integrated with the global classification loss function (i.e., the CE loss) in model training, as depicted in Fig. 1. The CE loss is used in this paper as follows. 
\begin{equation}
  L_{CE} = \frac{1}{\sum\limits_{n=1}^N|S_n|+|Q_n|}\sum\limits_{(\textbf{x},y) \in S \cup Q}-log p(y|\textbf{x};\theta)
\end{equation}
A hyperparameter $\lambda$ controls the balance of the CE loss with the PN loss.
\begin{equation}
  L_{CP}=L_{CE} + \lambda L_{PN}
\end{equation}

\vspace{-0.1cm}
\begin{figure}[htb]
  \centering
  \includegraphics[scale=0.85]{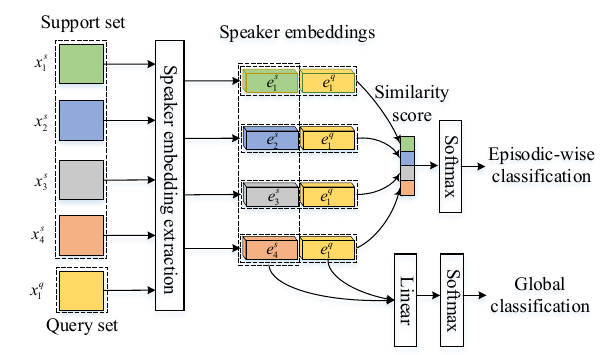}
  \caption{Meta-learning scheme with global classification.}
\end{figure}
\vspace{-0.3cm}

\subsection{Speaker embedding network}
A typical speaker embedding network contains several stacked frame-level layers, a statistics pooling layer and some utterance-level layers. We use the ResNet34 \cite{he2016deep} framework to extract frame-level representations in this paper, and a detailed configuration is listed in Table 1, where T denotes variable-length data frames. The input layer consists of a single convolutional layer with a kernel size of 7$\times$7, stride of 2$\times$2 and channel dimension of 32. Four residual stages include [3, 4, 6, 3] basic blocks with 32, 64, 128, and 256 channels respectively, and each basic block has 2 convolutional layers with filter sizes of 3$\times$3. Downsampling is performed by conv2\_1, conv3\_1, conv4\_1 and conv5\_1 with a stride of 1$\times$2.
\begin{table}[t]
\renewcommand\arraystretch{1.1}
\centering
\caption{The architecture of the speaker embedding network}
\begin{tabular}{cc|c|c|c}


\textbf{Layer} & \multicolumn{2}{|c|}{\textbf{Structure}} & \multicolumn{2}{c}{\textbf{Output size}}\\
\hline
\hline
Conv1 & \multicolumn{2}{|c|}{7 $\times$ 7, 64, stride 2} &  \multicolumn{2}{c}{$T$ $\times$ 40 $\times$ 32} \\
\hline
\multirow{2}{*}[1ex]{Conv2\_x} & \multicolumn{2}{|c|}{$\begin{bmatrix} 3\times3 & 32 \\ 3\times3 & 32 \\ \end{bmatrix}$ $\times$ 3} 
& \multicolumn{2}{c}{$T/2$ $\times$ 20 $\times$ 32} \\
\cline{1-5}
\multirow{2}{*}[1ex]{Conv3\_x} & \multicolumn{2}{|c|}{$\begin{bmatrix} 3\times3 & 64 \\ 3\times3 & 64 \\  \end{bmatrix} $ $\times$ 4} 
& \multicolumn{2}{c}{$T/2$ $\times$ 10 $\times$ 64} \\
\cline{1-5}
\multirow{2}{*}[1ex]{Conv4\_x} & \multicolumn{2}{|c|}{$\begin{bmatrix} 3\times3 & 128 \\ 3\times3 & 128 \\ \end{bmatrix}$ $\times$ 6} 
& \multicolumn{2}{c}{$T/2$ $\times$ 5 $\times$ 128} \\
\cline{1-5}
\multirow{2}{*}[1ex]{Conv5\_x} & \multicolumn{2}{|c|}{$\begin{bmatrix} 3\times3 & 256 \\ 3\times3 & 256 \\ \end{bmatrix}$ $\times$ 3} 
& \multicolumn{2}{c}{$T/2$ $\times$ 3 $\times$ 256} \\
\cline{1-5}
\multirow{2}*{} & \multicolumn{4}{|c}{\tabincell{c}{statistics pooling, 2 $\times$ \{512-d fc\}, softmax}} \\
\cline{1-5}
\end{tabular}
\vspace{-0.3cm}
\end{table}

After frame-level features are extracted from the above ResNet34 architecture, a statistics pooling layer converts the frame-level input to an utterance-level speaker representation. Then, two fully connected layers map the utterance-level features to speaker embeddings that are finally passed into a softmax output layer.

\section{Improved meta-learning training}

\subsection{ML with additional transformation coefficients}
The proposed Meta-Learning method with additional Transformation Coefficients (MLTC) is divided into two steps. In the first step, we train a backbone embedding network with a combination of the PN and CE loss functions; this step is similar to that described in \cite{kye2020meta}. In the following step, we append the pretrained backbone network (blue parts in Fig. 2) with some additional transformation coefficients (green parts in Fig. 2), and the additional transformation coefficients are trained with only the PN loss function with the parameters of the backbone network fixed.

\begin{figure}[htb]
  \centering
  \includegraphics[scale=0.9]{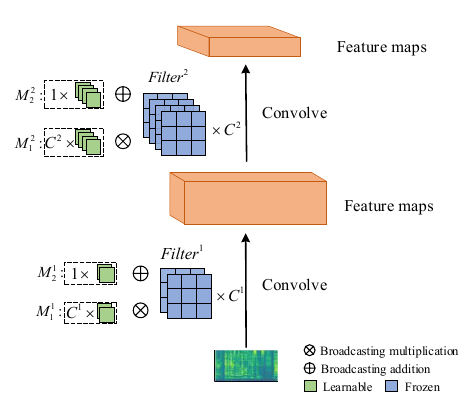}
  \caption{The proposed MTLC architecture.}
  \vspace{-0.3cm}
\end{figure}

Suppose the backbone network has well-trained parameters with filter weight matrices $\textbf{W}^{(l)}$ (for all $l$) after the first step, and the output of the $l^{th}$ layer can be computed as follows.
\begin{equation}
  \textbf{O}^{(l)}=relu((\textbf{W}^l \odot \textbf{M}_1^l)\textbf{O}^{(l-1)}+\textbf{M}_2^l)
\end{equation}
where $\odot$ denotes the broadcasting multiplication, $\textbf{O}^{(l)}$ denotes outputs from the $l^{th}$ layer and $relu(\cdot)$ stands for the ReLU activation function. $\textbf{M}_1^l$ and $\textbf{M}_2^l$ are two sets of transformation coefficients in the $l^{th}$ layer. The detailed algorithm is given as follows.

\begin{algorithm}[h]
	\caption{MLTC Training Procedure} 
	\begin{algorithmic}
		\REQUIRE Training data $S \cup Q$, learning rate $\alpha,\beta$
		\ENSURE Embedding extractor parameters $\theta$, transformation coefficients $M_1M_2$ \\
		Randomly initialize $\theta$; 
	\FOR{ samples in $S \cup Q$} 
		\STATE Calculate $L_{CP}$ by Eq. 5;
		\STATE Optimize $\theta$ with learning rate $\alpha$;
	\ENDFOR \\
	Fix $\theta$; \\
	Randomly initialize $M_1M_2$; \\
	\FOR{samples in $S \cup Q$} 
		\STATE Calculate $L_{PN}$ by Eq. 3; \\
		\STATE Optimize $M_1M_2$ with learning rate $\beta$; \\
	\ENDFOR
	\end{algorithmic}
\end{algorithm}
\vspace{-0.3cm}

\subsection{ML with the Addition Contrastive Loss}
In addition to the abovementioned PN and CE loss functions, the contrastive loss is used in the model training. Specifically, the random erasing (RE) data augmentation technique is applied on the support samples in each episode, and a small percent (approximately 10$\%$) of FBank features from the support sample $\textbf{x}_{n,i}^s$ are randomly set to zero. The augmented and original support samples $(\textbf{x}_{n,i}^s, \textbf{x}_{n,i}^{s_{aug}})$ can construct a new positive pair in this episode. The addition contrastive loss function for these newly constructed positive pairs is defined as follows.
\begin{equation}
\begin{split}
  L_{Contra} = &-\frac{1}{N}\sum\limits_{n=1}^N\frac{1}{S_n}\sum\limits_{(\textbf{x}_{n,i}^s,y_n^s) \in S_n} \\
  &log \frac{exp(-d(f_{\theta}(\textbf{x}_{n,i}^s), f_{\theta}(\textbf{x}_{n,i}^{S_{aug}})))}{\sum\limits_{k=1}^{N}exp(-d(f_{\theta}(\textbf{x}_{n,i}^s), f_{\theta}(\textbf{x}_{n,i}^{S_{aug}})))}
 \end{split}
\end{equation}
The speaker embedding network is jointly trained with the CE loss, PN loss and proposed ACL function. The overall loss is calculated.
\begin{equation}
  L_{CPC} = L_{CE} + \lambda(L_{PN} + L_{Contra})
\end{equation}
The specific algorithm is as follows.

\begin{algorithm}[h]
	\caption{ACL Algorithm} 
	\begin{algorithmic}
		\REQUIRE Training data $S \cup Q$, learning rate $\alpha$
		\ENSURE Embedding extractor parameters $\theta$ \\
		Randomly initialize $\theta$; 
	\FOR{support set $S=\{\textbf{x}_{n,i}^s\}$} 
		\STATE Apply RE data augmentation: $\textbf{x}_{n,i}^{S_{aug}}=RE(\textbf{x}_{n,i}^s)$;
		\STATE Construct positive pairs $(\textbf{x}_{n,i}^s, \textbf{x}_{n,i}^{S_{aug}})$;
	\ENDFOR
	\FOR{samples in $S \cup Q$} 
		\STATE Extract and normalize embeddings $\textbf{z}_{n,i}^s, \textbf{z}_{n,i}^q, \textbf{z}_{n,i}^{S_{aug}}$;
		\STATE Calculate $L_{Contra}$ by Eq. 7 and $L_{CPC}$ by Eq. 8;
		\STATE Optimize $\theta$ with learning rate $\alpha$;
	\ENDFOR
	\end{algorithmic}
\end{algorithm}
\vspace{-0.3cm}

\section{Experiments and analysis}
\subsection{Experimental settings}
\subsubsection{Datasets and evaluation metrics}

To investigate the effectiveness of the proposed methods, we conduct experiments on the SITW and VOiCES datasets. The development portions of VoxCeleb1 \cite{nagrani2017voxceleb} and VoxCeleb2 \cite{chung2018voxceleb2} are used for training. There are 60 speakers included in both SITW and the development portion of VoxCeleb1. These speakers are removed from the training dataset.

For the SITW dataset \cite{mclaren2016speakers}, there are two standard sets for testing: dev. core and eval. core. We use both sets for experiments. The VOiCES dataset for the speaker verification task is described in "VOiCES from a Distance Challenge 2019" \cite{nandwana2019voices}. The VOiCES development consists of 15,904 audio segments from 196 speakers and includes 20,224 target and 4,018,432 impostor trials. The evaluation set consists of 11,392 audio segments from 100 speakers and includes 36,443 target and 357,073 impostor trials. Due to the background noise, reverberation and laughter contained in the speech data, the data augmentation techniques described in \cite{snyder2018x}, including adding additive noise and reverberation data, are applied to improve the robustness of the system.

The results are reported in terms of two metrics, namely, the EER and the minimum of the normalized detection cost function (minDCF) with the prior target probability $P_{tar}$ set to 0.01.

\subsubsection{Input features}

The feature extraction process uses the Kaldi toolkit \cite{povey2011kaldi}. In our implementation, audio signals are first transformed into 25ms width frames with 10ms frame shifts. Then, we select the 40-dimensional FBank features as the input acoustic features. Energy-based voice activity detection (VAD) is used to remove nonspeech frames. The acoustic features are randomly truncated into short slices ranging from 2 to 4 s.

\subsubsection{Speaker embedding extraction networks}

For comparison, five systems with different training strategies are listed as follows. All systems have the same ResNet34 architecture described in Table 1. The neural networks are trained with the Tensorflow toolkit \cite{abadi2016tensorflow} and optimized with the Adam optimizer. The embeddings are extracted from the first fully connected layer with dimensions of 512. The same type of batch normalization and L2 weight decay described in \cite{zeinali2019improve} are used to prevent overfitting.
In the training process, the minibatch size is set to 80. We randomly sample 20 speakers with 4 utterances in each speaker from the training set and then sample 1 and 3 utterances from each speaker as the support set and query set, respectively. Other configurations of systems are listed as follows.

\begin{table*}[t]
    \vspace{-1cm}
    \caption{Results of different systems on the SITW and VOiCES datasets}
	\centering
	\begin{tabular}{cp{1cm}p{1cm}p{1cm}p{1cm}p{1cm}p{1cm}p{1cm}p{1cm}}
	\toprule
        \multirow{4}*{Systems} & \multicolumn{4}{c}{SITW} & \multicolumn{4}{c}{VOiCES} \\
        \cmidrule(l){2-5} \cmidrule(l){6-9}
		 \multirow{1}*{} & \multicolumn{2}{c}{Dev} & \multicolumn{2}{c}{Eval} & \multicolumn{2}{c}{Dev} & \multicolumn{2}{c}{Eval} \\
		 \cmidrule(l){2-3} \cmidrule(l){4-5} \cmidrule(l){6-7} \cmidrule(l){8-9}
		 \multirow{1}*{} & EER & DCF & EER & DCF & EER & DCF & EER & DCF 	\\
        \midrule    
        Baseline & 1.694 & 0.2159 & 1.914 & 0.2396 & 1.891 & 0.1949 & 6.048 & 0.4329 \\
		 PN & 1.502 & 0.1984 & 1.751 & 0.2326 & 1.667 & 0.2096 & 5.812 & 0.4291 \\
        MLTC & 1.386 & 0.1865 & 1.640 & 0.2197 & 1.443 & 0.1856 & 5.518 & 0.3975 \\
		 ACL & 1.388 & 0.1812 & 1.531 & 0.2103 & 1.468 & 0.1867 & 5.483 & 0.3981 \\
		 \midrule 
		 MLTC\&ACL & 1.348 & 0.1793 & 1.504 & 0.2035 & 1.423 & 0.1762 & 5.332 & 0.3911 \\
		 Score fusion & \textbf{1.309} & \textbf{0.1732} & \textbf{1.476} & \textbf{0.2008} & \textbf{1.289} & \textbf{0.1577} &      \textbf{5.096} & \textbf{0.3642} \\
	 \bottomrule
    \end{tabular}
    \vspace{-0.3cm}

\end{table*}

\noindent 
\textbf{Baseline}: This is the conventional ResNet34 architecture trained with the CE loss. The learning rate gradually decreases from 1e-3 to 1e-4 in the training process.

\noindent 
\textbf{PN system}: The network in Section 2.1 is trained with the loss function defined in Eq. 5, where the parameter $\lambda$ is set to 0.5 and the other setup is the same as that of the baseline.

\noindent 
\textbf{MLTC system}: The network in Section 3.1 is trained in two steps. In the first step, the abovementioned PN system is trained. In the second step, the additional transformation coefficients are trained with the same training set, and the learning rate gradually decreases from 1e-4 to 1e-5.

\noindent 
\textbf{ACL system}: The RE data augmentation strategy described in Section 3.2 is employed, and the network is trained with the loss function defined in Eq. 8, where the parameter $\lambda$ is also set to 0.5 and the other configuration is the same as that of the baseline.

\noindent 
\textbf{MLTC\&ACL system}: Both the MLTC and ACL methods are employed in this system. First, the RE data augmentation technique is applied on the support set and more positive pairs are constructed. Then, the backbone network is trained with $L_{CPC}$ in Eq. 8. These two steps are same as the steps in the above ACL system. When the backbone network converges, the additional transformation coefficients $S_1S_2$ in Eq. 6 are optimized with the PN loss, which is similar to the process in  MLTC system. The learning rate gradually decreases from 1e-4 to 1e-5 in this step.

\noindent 
\textbf{Score level fusion}: The complementarity between the MLTC system and ACL system at the score level is also investigated here. We only report the results with the score-level fusion of the two systems with equal weights.

\subsubsection{Backend algorithm}

The embeddings are centered with the training set and are projected onto a low-dimensional space with LDA at first. The dimension of the speaker embedding is reduced to 120. After length normalization, we select the longest 200,000 recordings from the training set to train the PLDA backend.

\subsection{Results and analysis}

Table 2 presents the results of different systems on the SITW and VOiCES datasets. It can be observed that the PN system outperforms the conventional ResNet34 baseline. This finding verifies the effectiveness of the fusion of meta-learning scheme and global classification.

Moreover, both the MLTC and ACL systems can achieve improvements over the PN system by approximately 10$\%$ in terms of minDCF and EER. As a single system, the MLTC$\&$ ACL system can outperform the MLTC and ACL systems on all the evaluation conditions. Furthermore, the score-level fused system in the last row achieves the best results. The results on the last two rows demonstrate that these two improved methods are highly complementary.

\subsection{Comparison with conventional fine-tuning}

In this section, we conduct a toy experiment to demonstrate the necessity of the additional transformation coefficients for prototypical networks training in Section 3.1. We do not use the additional transformation coefficients and directly fine-tune the backbone network with Eq. 3, denoted as MLFT. The experimental results are shown in Table 3. The performance of the MLFT system is far worse than that of MLTC or even the PN systems. This demonstrates that fine-tuning the backbone network with only the PN loss may change the original pattern of network. Not only can it not bring performance gains, but it may degrade performance.

\vspace{-0.2cm}
\begin{table}[h]
    \caption{Comparison results of MLTC and MLFT}
    \centering
	\begin{tabular}{cp{1cm}p{1cm}p{1cm}p{1cm}}
	\toprule
        systems & \multicolumn{2}{c}{SITW(Eval)} & \multicolumn{2}{c}{VOiCES(Eval)} \\
        \cmidrule(l){2-5}
		\multirow{2}*{} & EER & DCF  & EER & DCF  	\\
    \midrule     
        PN & 1.751 & 0.2326 & 5.812 & 0.4291  \\
		MLTC & \textbf{1.640} & \textbf{0.2197} & \textbf{5.518} & \textbf{0.3975} \\
		MLFT & 1.804 & 0.2359 & 5.856 & 0.4383 \\
	 \bottomrule
    \end{tabular}
    \vspace{-0.3cm}

\end{table}

\subsection{Comparison with data augmentation on the query set}

In the ACL system, we apply the RE data augmentation technique on the support samples. In this section, the same operation is applied over the query samples, denoted as ACL-Q.  All other operations are the same as those described in Section 3.2. The experimental results are shown in Table 4, where ACL-Q stands for the contrastive system.

\vspace{-0.2cm}
\begin{table}[h]
    \caption{Comparison results of ACL and ACL-Q}
    \centering
	\begin{tabular}{cp{1cm}p{1cm}p{1cm}p{1cm}}
	\toprule
        systems & \multicolumn{2}{c}{SITW(Eval)} & \multicolumn{2}{c}{VOiCES(Eval)} \\
        \cmidrule(l){2-5}
		\multirow{2}*{} & EER & DCF  & EER & DCF  	\\
    \midrule     
        PN & 1.751 & 0.2326 & 5.812 & 0.4291  \\
		ACL & \textbf{1.531} & \textbf{0.2103} & \textbf{5.483} & \textbf{0.3981} \\
		ACL-Q & 1.777 & 0.2280 & 5.782 & 0.4380 \\
	 \bottomrule
    \end{tabular}
    \vspace{-0.3cm}

\end{table}

It can be seen that the RE data augmentation technique on the query samples cannot yield benefits. Since various data augmentation methods have been applied to the original audio described in Section 4.1.1, extra data augmentation on the query samples can only achieve marginal improvements. The improvement obtained by the ACL system stems from the application of contrastive learning ideas.

\section{Conclusions}

In this paper, we propose the MLTC and ACL methods to improve meta-learning scheme. Both of these methods can achieve consistent improvements over the conventional prototypical network on the SITW and VOiCES databases. Furthermore, these two methods can be integrated into one framework to provide extra improvements.

\section{Acknowledgements}
This work was partially funded by the National Natural Science Foundation of China (Grant No. U1836219).

\bibliographystyle{IEEEtran}

\bibliography{mybib}


\end{document}